\documentclass{jsara}
\journalinfo{Journal of the Southeastern Association for Research in Astronomy, {\bf 7}, 56-59, 2012 December 1}

\slugcomment{}

\shorttitle{\textsc{CMD of M14}}
\shortauthors{\textsc{Reinhart et al.}}

\begin{document}

\setcounter{page}{00}


\title{The Color-Magnitude Diagram of the Globular Cluster M14\altaffilmark{1}}


\author{Erik D. Reinhart\altaffilmark{2,3}}
\affil{Department of Physics, Willamette University, Salem, Oregon, 97301}

\author{Thayne McCombs\altaffilmark{2,3}}
\affil{Department of Physics and Astronomy, Brigham-Young University-Provo, Provo, Utah, 84602}

\author{Andrew N. Darragh, Zheyu J. Liu, Brian W. Murphy}
\affil{Department of Physics and Astronomy, Butler University, Indianapolis, Indiana 46208}

\and

\author{Kyle E. Conroy\altaffilmark{2,3}}
\affil{Department of Astronomy and Astrophysics, Villanova University, Villanova, PA, 19085}

\altaffiltext{1}{Based on observations obtained with the SARA
  Observatory 0.9m telescope at Kitt Peak national Observatory, which is owned and
  operated by the Southeastern Association for Research in Astronomy
  (http://www.saraobservatory.org). }

\altaffiltext{2}{Southeastern Association for Research in Astronomy
(SARA) NSF-REU Summer Intern}
\altaffiltext{3}{Visiting observer, SARA Observatory at Kitt Peak, Arizona, which is operated by
the Southeastern Association for Research in Astronomy}

\email{ereinhar@willamette.edu, bmurphy@butler.edu}


\newcommand{\vdag}{(v)^\dagger}
\newcommand{\myemail}{ereinhar@willamette.edu}


\begin{abstract}
Using the SARA 0.9 meter telescope at Kitt Peak National Observatory (KPNO) we obtained $R$ images during the summer of 2010 and $V$  images during the summer of 2012 of the globular cluster M14. These images were analyzed using the DAOPHOT/ALLFRAME package of \citet{stetson87,stetson94} to create a preliminary color-magnitude diagram (CMD) of M14.   We have positively identified the positions of 64 of the 112 known RR Lyrae stars on the CMD. We find considerable spread in both the magnitudes and color of the RR Lyrae stars indicating a significant amount of differential reddening on relatively small scales.  This is also seen in the giant branch which shows up to a $\sim$0.15 magnitude spread in $V-R$ color.  We also found the median magnitude of the RR Lyrae stars and hence the horizontal branch to be $V$=17.2. \end{abstract}


\keywords{clusters: globular, stars: variables: general-Galaxy: individual: M14}


\section{Introduction}

Globular clusters are relatively old and concentrated star clusters found in the galactic halo. Because the stars in a given cluster are presumed to have formed from the same cloud of gas and dust  they most likely have similar composition, age, and distance. This makes globular clusters ideal laboratories for studying the evolution of a single population of stars.  Depending on their metallicity, age, and mass globular clusters can have large numbers  of RR Lyrae stars.  RR Lyrae stars are pulsating variable stars located on the horizontal branch with approximate absolute magnitudes of $V$=0.5, making them ideal standard candles. They come in two general types, RR0 variables which have longer periods and larger amplitudes than RR1 variables which have a more sinusoidal light curve.  Though they have very similar $V$ absolute magnitudes RR1 variables are slightly bluer than their RR0 counterparts, differing in color by roughly 0.1 in $(V-R)$.  By using Fourier decomposition of the variable light curves it is possible to compare the amplitudes and phases of the light curves to hydrodynamic stellar evolution models.  From these comparisons estimates of the metallicity, helium abundances, and ages of these stars can be made \citep{simon82,simon93}.

Early studies of variable stars in globular clusters were made using photographic plates, and despite the errors involved in their use many variables were found, the most common being of RR Lyrae type \citep{wehlau94}.  However, even to this day the cores of many globular clusters are yet to be examined using modern techniques to search for variables.  With the use of CCDs and image subtraction software it is possible to discover heretofore unknown variable stars \citep{alard98,alard00}.   In many cases the number of known variables in a cluster may double using these new techniques.   Having a larger sample of variables can allow for a more detailed analysis of a cluster and a better determination of its physical parameters.

Previous work by our research group has focused on several globular clusters, M14 (NGC~6402), NGC~4833, NGC~6584, and more recently M107 (NGC~6171) \citep{conroy12,darragh12,toddy12,mccombs12}.  In this study we revisit M14.  Recently \citet{conroy12}, using the image subtraction method, found 59 previously unknown RR Lyrae variables in M14, more than doubling the total and raising the number to 112.     Because of this relatively large population of  RR Lyrae variables M14 became an ideal case for further study.

M14 is relatively close to the galactic center ($l=21.3^{\circ}$, $b=14.8^{\circ}$).  Because of this proximity M14 is significantly reddened due to interstellar dust and has a listed $E(B-V$) of 0.6 implying nearly 2 magnitudes of extinction in $V$  \citep{harris96}.   It is also likely that this extinction is not uniform but instead patchy across the cluster.   Figure 1 shows a $5\times5^{\circ}$ region at 100 microns centered on M14.  It is clear that there is significant differential reddening across the region.  This differential reddening can prevent an accurate determination of the distance, age, and metallicity of the cluster.  It is also a concern when constructing a CMD of the cluster.
In spite  of this difficulty we have used data taken with the Southeastern Association for Research in Astronomy (SARA)  telescopes to  create a CMD and identify the location of RR Lyrae stars on it.    From  these CMD positions we are then able to  verify  if they are consistent with cluster membership.  The CMD  also  allows us  to estimate the amount of differential reddening in the vicinity of M14.
\begin{figure}
\epsscale{1.18}
  \plotone{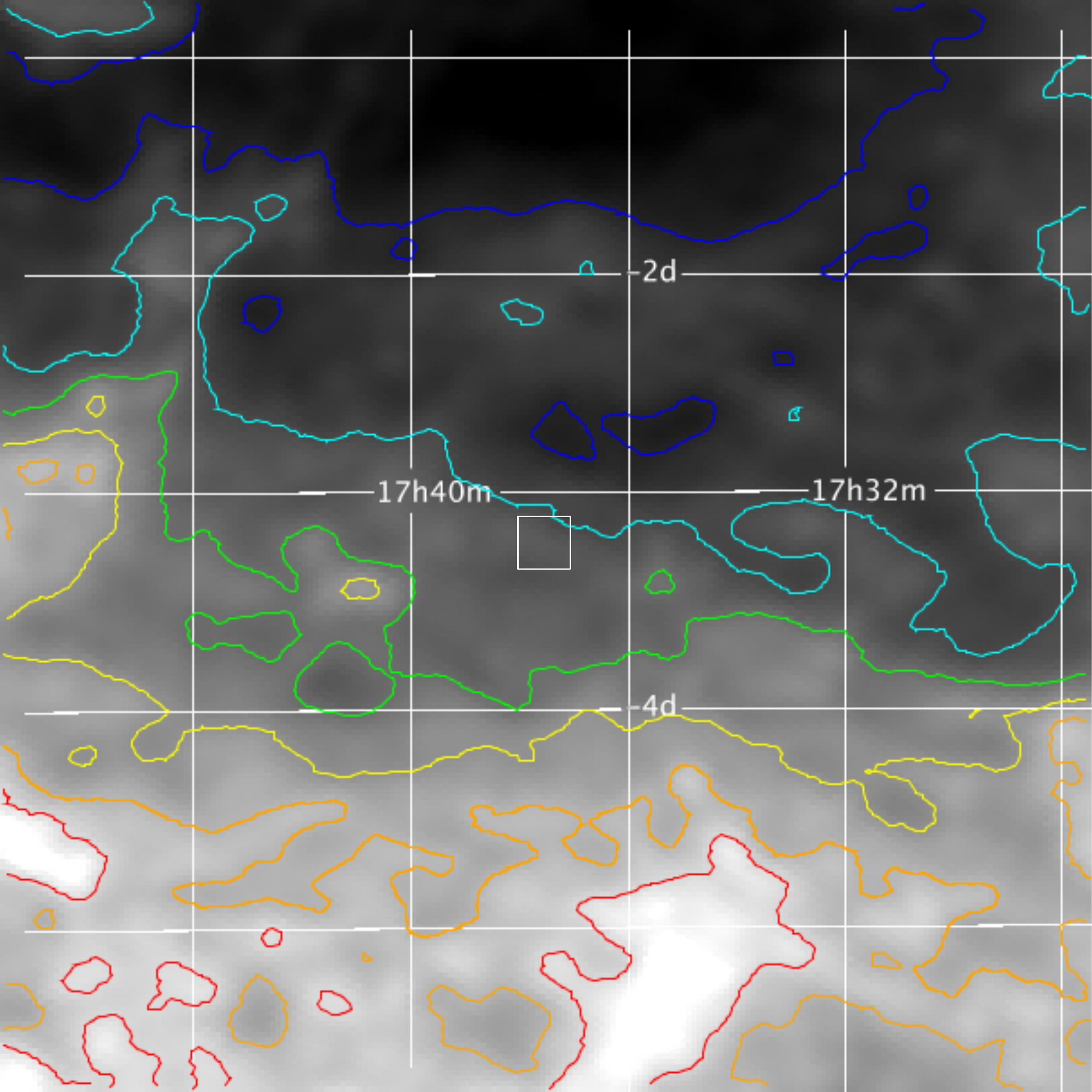}
\caption{100 micron emission in the vicinity of the globular cluster M14 (Schlegel, Finkbeiner, \& Davis 1998).  This is a reprocessed composite of the COBE/DIRBE and IRAS/ ISSA maps, with the point sources removed.  The contours cover the range from  9.01 (blue) to 22.31 MJy/sr (red) in intervals of 2.22 MJy/sr.  The square near the center of this image represents our field of view with SARA KPNO.}
\label{fig1}
\end{figure}

\section{Observations \& Reduction}
$R$-band image frames used in this study were published by \citet{conroy12}.  Those observations were obtained in May and June of 2010 using the KPNO SARA 0.9 meter telescope with an Apogee Alta U42 CCD with a 2048$\times$2048 Kodak e2V CC42-40 chip, a gain of 1.2 electrons per count, RMS noise of 6.3 electrons, and cooled to a temperature of approximately $-30^{\circ}{\rm C}$. 1$\times$1 binning was used, resulting in a scale of 0.42\arcsec/px and a 13.6$\times$13.6\arcmin\ field of view. Typical seeing was 2.2\arcsec\ and ranged from 1.5-3.0\arcsec.  Exposure times were set at a constant 60 seconds to avoid overexposure of the core or bright giants during best possible seeing conditions. $V$-band images were obtained during the summer of 2012 also using the SARA 0.9 meter telescope on the nights of  14 May and 7, 8, 9 June 2012. The same Apogee Alta U42 CCD was used with an exposure time of 100 seconds,  however we used 2$\times$2 binning. This resulted in a scale of 0.825\arcsec/px and a 13.6$\times$13.6\arcmin\ field of view.  We used standard data reduction procedures using the Maxim DL image processing software.  Because hot or dead pixels as well as cosmic rays can produce a large number of false positives when running the image subtraction software ISIS, we also applied a bad pixel map and ran Maxim DL's hot and dead pixel kernel filters.  We then applied  the {\tt CRAVERAGE} cosmic ray removal utility in IRAF to each frame.









\begin{figure*}
  \epsscale{1.}
  \plotone{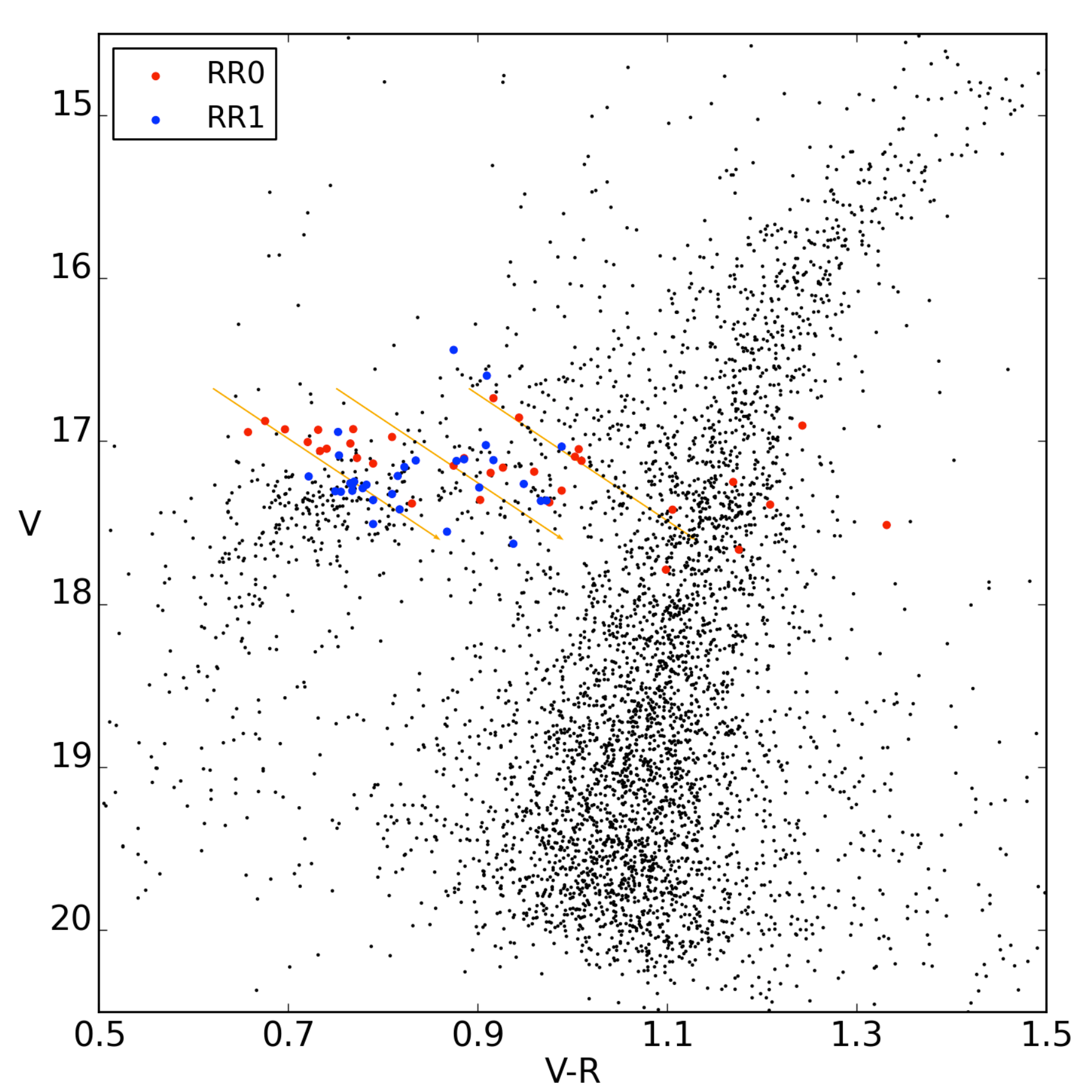}
\caption{Color Magnitude Diagram of M14 from two $V$ and R.  464 images were stacked together to  produce the $V$  image and 698 images were stacked to produce the $R$ image.  The blue points are the RR1 stars and the red are the RR0.  Much of the scatter in the RR lyre stars is due the result of stacking images to improve signal to noise and due to differential reddening.  The diagonal lines indicate the reddening tracks.  }
\label{fig2}
\end{figure*}
\section{Analysis}
\subsection{Image Subtraction} \label{bozomath}
 ISIS v2.2 was used to search for, identify, and determine the positions of the variables (Allard
\& Lupton 1998; Allard 2000). The package contains seven c-shell scripts and three parameter files that were used in our analysis.  ISIS uses pattern recognition to align stars from image to image and search for changes in brightness of every star it finds.  ISIS creates a reference image from a set of the highest quality images so that all other images are compared/subtracted from this.  ISIS also takes into account differences in seeing. To do this ISIS convolves the reference image so that the point-spread function of the stars is nearly identical to that of the image it is being subtracted from. Note that ISIS uses relative flux not magnitudes. The ISIS parameters and methods that we used are described in more detail in the analysis of M14 by \citet{conroy12}.

\subsection{Photometry} \label{bozomath}
While ISIS very useful for searching for variables it does not allow for absolute photometry of the variables.  The relative flux output from ISIS is not adequate since the magnitude scale is not linear.  The relative fluxes are useful for determining period and variable type but of little use in determining most photometric properties of variables. Therefore it is necessary to obtain instrumental magnitudes of each of the variables found in addition to other stars in the cluster so that a CMD can be created and cluster membership confirmed.   Another added difficulty are the crowded fields typically found in globular clusters, particularly their cores.  To deal with the crowded nature of globular clusters it has become standard procedure to use Stetson's DAOPHOT/ALLFRAME \citep{stetson87,stetson94} to create CMDs.  The combination of these packages results in magnitudes for both $V$  and $R$ that we then  could match using a script written by \citet{tabur07}.

\subsection{Magnitude Transformation} \label{bozomath}
ALLFRAME gives instrumental magnitudes of each star in the images with an arbitrary zero point of 25 and does not take into account exposure time, extinction, or color. We transformed the instrumental magnitude $V_{inst}$ to true apparent magnitude $V_{app}$ by using a set of standard stars within the cluster as identified and calibrated by \citet{stetson00}.  These standard stars covered a wide range of magnitudes that resulted in a linear relationship between our instrumental magnitude and Stetson's standards.  The next step was to transform the instrumental  $(V-R)$ to true values.   This was done using transformation coefficients determined earlier in 2012 by one of the authors.
It is also worth mentioning that because of the differing binning between the $V$  and $R$ images the $R$ images were converted from 1$\times$1 to 2$\times$2 so that proper alignment and colors could be determined.  This resulted in counts that were 4 times less than would be expected.  The factor of 4 difference occurs when MaxIm DL does pixel operations of this nature.  Maxim DL  interpolates over the pixels and point spread functions  then assigns a newly binned pixel value that is equal to the median of the original pixels.  In reality the new pixel value should be proportional to the increase in area of the new $2\times2$ rebinned pixel, i.e.,  a factor of 4.  To correct for this we multiplied each pixel in the $R$ frame by 4.

\section{Results} \label{bozomath}





The color magnitude diagram of M14 was created using $V$  and $R$ band KPNO data. The $V$  and $R$ average images were created by stacking images. The $V$ stacked image was an average of 464 images and the $R$ stacked image was created using an average of 698 images.  Because of this stacking the magnitudes of our RR Lyrae variables are not average values but dependent upon when they were observed.   This results in some scatter of these variable stars in the CMD.  Stars in the $V$ and $R$  stacked images were matched using a script  \citep{tabur07}.  The script creates a transform between the coordinates of the stars detected in the  $V$ and $R$ stacked frames, and only retains those stars that are detected in both colors.  Thus any stars that did not have matches in each color were eliminated by the script.  This process resulted in 4721 star matches in both $V$ and $R$ that were then be plotted in the CMD.

ISIS and ALLFRAME both give positions of variables and all stars found, respectively.  But because ISIS used only a few select good seeing frames that were combined into a reference image for subtraction and ALLFRAME used a much larger number of stacked frames, the resulting positions of the variables found by ISIS can vary somewhat from ALLFRAME.  To ensure that we were properly pulling our variables out of ALLFRAME we only plotted variables found by ISIS that were within 0.5\arcsec\  of the ALLFRAME position.  That lowered the scatter of the variables on the CMD and eliminated any misidentifications.  In total we were able to identify and plot 33 RR0 and 31 RR1 variables onto the CMD.

The CMD for M14 is shown in Figure 2.   The RR Lyrae stars are shown by the colored dots, with the RR0's being red and the RR1's shown in blue.  The diagonal lines indicate the direction of reddening.   It is apparent that there is significant scatter in the diagram.  Most of scatter is due to differential reddening.   Compared to other clusters observed by this group the scattered found here is much greater than say for NGC~6584 which is much less reddened than M14 \citep{toddy12}.   As expected the largest scatter is found among the RR0 variables since they have the largest amplitudes and greatest change in color from minimum to maximum phase.  The scatter is somewhat less for the RR1 variables due to  their smaller amplitudes.  Given their shorter periods of roughly 8 hours we were more likely to be close to their time-averaged magnitudes when stacking images to produce the CMD.  The two RR1 variables found well above the horizontal branch are likely blended with redder stars.  Several of the RR0 variables to the upper left of the instability strip were probably those that happened to be imaged each night near maximum.  At peak brightness RR0 variables are much bluer than average, thus this scatter can mimic reddening. We found the median $V$  magnitudes and $(V-R)$ color of the RR0 variables to be 17.26 and 0.91, respectively.  For the RR1 variables we found these quantities to be 17.10  and 0.82, respectively.   The color difference is typical given that RR1 variables are warmer than their RR0 counterparts.  Using all 64 of these RR Lyrae variables we find a horizontal branch $V$  magnitude of approximately 17.17, similar to but slightly brighter than the cataloged value of 17.3 \citep{harris96}.

By examining the giant branch of the cluster it is possible to estimate the degree of differential reddening.  The $\Delta(V-R)$ is roughly 0.1 to 0.15 from one side of the giant branch to the other.  This is greater than the expected photometric errors.  Given this spread of 0.15 magnitudes in color the expected $\Delta{V}$ could be 0.5 to 0.75 magnitudes.  This is close to what is seen in the $V$  variation among the RR Lyrae variables, indicating significant differential reddening as would have been expected from the 100 micron image shown Figure 1.

\section{Conclusions}
Due to time constraints, we have only presented a preliminary analysis of the the CMD and location of the RR Lyrae variables in M14.  From the 64 positively identified variables on the CMD we estimate a horizontal branch $V$  magnitude of 17.2.  Our analysis has shown there is significant scatter in the RR Lyrae variables, much of it due to differential reddening that must be addressed in our future investigations in order to determine a more accurate distance, age and metallicity of the cluster M14.  To accomplish this our next task will be to determine the time-weighted $V$  and $R$ magnitudes for the RR Lyrae variables via full Fourier decomposition. Once this is completed we will be better able to determine physical parameters of the RR Lyrae variables and those of M14.




\acknowledgments

We thank C. Alard for making ISIS 2.2 publicly avail-
able and P Stetson for useful advise in using DAOPHOT and ALLFRAME. This project
was funded in part by the National Science Foundation Research Experiences for
Undergraduates (REU) program through grant NSF AST-1004872. Additionally A.
Darragh, Z. Liu, and B. Murphy were partially funded by the Butler Institute
for Research and Scholarship. The authors also thank F. Levinson for a generous
gift enabling Butler University's membership in the SARA consortium.

\clearpage


\clearpage


\clearpage

\clearpage




\end{document}